# Lingering Sweetness of Ethanol Clusters: Sensory Discovery and Objective Discrimination by Impedance-Based Electronic Tongue

Jiaxin Peng[1], Chaoyu Zhao[1], Xinyue Jiang[2,3], Yuqun Xie[1,3*]

[1] School of Life and Health Sciences, Hubei University of Technology, Nanli Road 28, Hongshan District, Wuhan 430068, China

[2] School of Civil, Architectural and Environmental Engineering, Hubei University of Technology, Nanli Road 28, Hongshan District, Wuhan 430068, China

[3] Key Laboratory of Intelligent Health Perception and Ecological Restoration of River and Lake, Ministry of Education, Wuhan 430068, China

[*] Email: yuqunxie@hbut.edu.cn (Y. X.)

**Abstract**

Ethanol is conventionally perceived only as a pungent tastant, while the potential sweet properties of ethanol clusters have remained unrecognized. Here we show that ethanol tetramers exhibit a unique time-dependent lingering sweetness, distinct from the immediate upupfront sweetness of conventional sweeteners. To objectively verify this phenomenon, we developed an impedance-based bionic electronic tongue that monitors changes at a lipid polymer membrane interface. Sensory evaluation revealed increasing ethanol cluster content from 4.83% to 29.53% prolonged sweetness duration from ~10 s to ~25 s ($p < 0.001$), whereas sweetness intensity remained low to moderate (1.0–1.7 on a 10-point scale). Using this electronic tongue, we achieved objective discrimination between upfront and lingering sweetness: full-band phase-angle scanning identified optimal frequencies of 0.933 Hz ($\tau \approx 0.171$ s) for xylitol (fast binding and dissociation) and 0.03 Hz ($\tau \approx 5.305$ s) for ethanol clusters (slow binding and dissociation). Phase-angle responses correlated strongly with sensory sweetness duration ($R^2 = 0.941$ and 0.931, respectively). In natural aged Chinese Baijiu (1–20 years), low-frequency phase-angle signals positively correlated with sensory lingering sweetness scores, confirming that ethanol cluster accumulation during aging underpins the sweetness of Baijiu. This work expands the classic AH-B sweetness theory from single molecules to supramolecular clusters and provides an analytical platform for probing the temporal dimension of flavor perception.

## 1. Introduction

From the conventional perspective, ethanol is a flavor substance with pungent odor and irritating taste. Yet, anyone who has tasted a well-aged spirit whether Chinese Baijiu, whiskey, or brandy knows that prolonged aging dramatically transforms this harsh profile into a smoother, more rounded mouthfeel(Jiang et al., 2026), often accompanied by a distinct and lingering sweetness(Jiang, Liu & Xie, 2024). This remarkable sensory improvement occurs without any significant increase in conventional sweeteners, leaving the molecular origin of this aging-induced sweetness a long-standing mystery. The classic AH-B glucophore theory(Zheng, Xu, Fang, Sun & Liu, 2024), proposed by Shallenberger and Acree, stipulates that a sweet-tasting molecule must possess a hydrogen-bond donor (AH) and acceptor (B) separated by 2.5–4.0 Å to activate the sweet taste receptor. As shown in Figure 1a,the cyclic ethanol tetramer a supramolecular cluster formed by four ethanol molecules linked via intermolecular hydrogen bonds exhibits an AH-B distance that falls exactly within this critical range. This structural coincidence raises the possibility that ethanol clusters, rather than free ethanol, may act as genuine sweet entities, providing a plausible explanation for the emergence of lingering sweetness in aged alcoholic beverages.

The AH-B theory, however, only answers whether a molecule is sweet, not how it elicits sweetness over time. Recent studies on sweet taste receptor dynamics (TAS1R2/TAS1R3) (Belloir, Brulé, Tornier, Neiers & Briand, 2021; Juen et al., 2025; Lu, Ma, Meng & Cui, 2025), reveal that the binding and dissociation rates(Servant & Kenakin, 2024) govern the temporal profile of sweetness. Within this framework, upupfront sweetness(Roelse, Krasteva, Pawlizak, Mai & Jongsma, 2024) refers to the immediate sweet sensation that appears within 0–3 s after stimulation and fades rapidly, corresponding to fast binding and dissociation. Lingering sweetness appears with a delay after stimulus removal and persists for a long time, corresponding to slow binding and dissociation(Deng et al., 2023; Yuan et al., 2024). Accurately characterizing these two sweetness modalities requires analytical tools that can capture dynamic molecular interactions. Sensory evaluation, though the gold standard, suffers from subjectivity(Yang, Cao, Qian & Chen, 2026) and cannot reliably quantify(Deng et al., 2023) lingering sweetness(Yuan et al., 2025). Chromatography identifies chemical composition but not dynamic receptor interactions(Zhang, Lin, Wang & Li, 2025). Bionic electronic tongues(Jeong et al., 2022; Liang, Zhou, Zhang, Xiao & Wu, 2026) have emerged as promising alternatives(Jung et al., 2023). The open-circuit-potential electronic tongue developed by Toko's group(Toko, 2023) successfully detects charged tastants by mimicking lipid membrane potential changes(Wu & Toko, 2023; Wu, Tahara, Yatabe & Toko, 2020). However, it faces severe sensitivity limitations with neutral molecules like ethanol, which hardly induce charge transfer at the lipid interface. More critically, existing sweetness electronic tongues， whether potentiometric or other types output only steady-state response values(Zhang, Wang, Huang, Liu & Tan, 2019) and cannot capture the binding and dissociation kinetics(Deng et al., 2023) that distinguish upupfront sweetness from lingering sweetness. This methodological gap has remained unaddressed for years: no electronic tongue has been reported to separate the temporal stages of sweet perception.

To fill this gap, this study aims to answer two core questions. First, do ethanol clusters possess independent sweet taste activity. Hypothesize that these supramolecular structures confer a unique time-dependent lingering sweetness, fundamentally different from the immediate upupfront sweetness of conventional sweeteners like xylitol. Second, can an electronic tongue objectively discriminate these two sweetness modalities.As shown in Figure 1cconvert the signal acquisition mode from open-circuit potential to electrochemical EIS (EIS), which monitors changes in the dielectric constant and charge transfer resistance at the lipid and polymer membrane interface upon sweet molecule binding. This allows, for the first time, the distinction and kinetic characterization of upfront vs. lingering sweetness based on their characteristic frequency responses and relaxation times. The findings are validated on real Chinese Baijiu samples aged from 1 to 20 years, linking low-frequency phase-angle signals to sensory lingering sweetness scores. This work expands the classical AH-B theory from single molecules to supramolecular clusters and provides an innovative analytical platform for probing the temporal dimension of flavor perception.

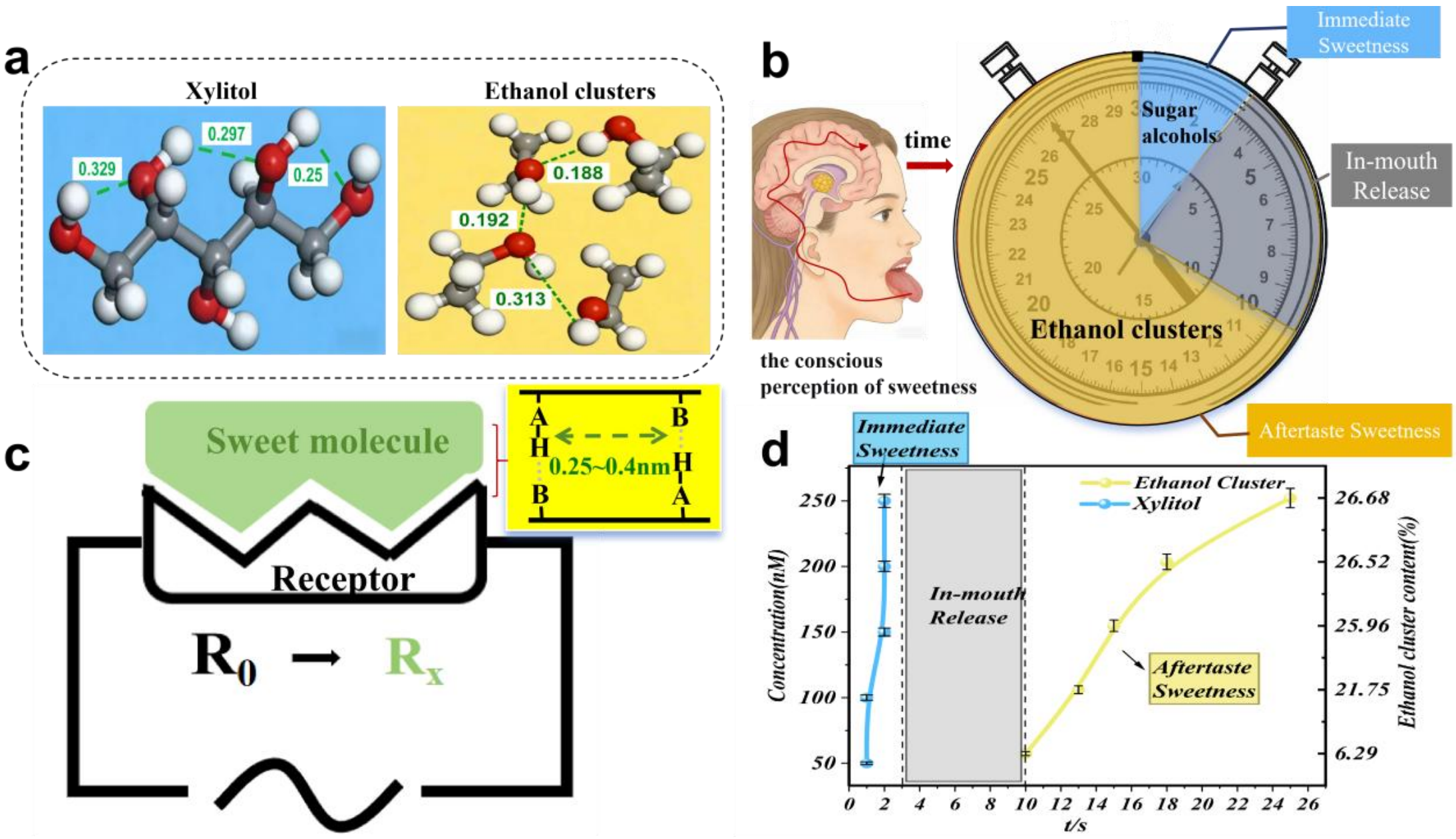

Figure 1 Differences in sweet taste mechanism and perception kinetics between xylitol and ethanol clusters. (a) Schematic diagram of molecular structures and hydrogen bond lengths of xylitol and ethanol clusters; (b) Schematic diagram of human temporal profile of sweet taste perception; (c) Schematic diagram of impedance-based sweet taste detection improved based on electronic tongue principle; (d) Comparison of sensory sweet taste duration between xylitol and ethanol clusters.

## 2. Materials and Methods

### 2.1 Experimental materials

Reagents: Xylitol, anhydrous ethanol (analytical grade, ≥99.7%), ultrapure water (resistivity ≥18.2 MΩ·cm), anhydrous sodium sulfate (analytical grade, ≥99.0%), tetraalkylammonium bromide (≥98%), di-n-octylphenyl phosphonate (≥97%, free of aromatic ring impurities), polyvinyl chloride (purity ≥99%, 80-120 mesh), trimellitic acid (≥98%), tetrahydrofuran (≥99.8%), potassium chloride, tartaric acid, sodium hydroxide, sodium chloride, sucrose, xylitol, alcohol dehydrogenase (ADH), nicotinamide adenine dinucleotide (NAD+). Strong-aroma type Chinese liquors were purchased from Huangshantou Group Co., Ltd. (Hubei, China). Clay pot samples naturally aged for 1, 4, 7, 10, 13 and 20 years were collected. The alcohol content of all samples was 69% (volume ratio).

Sample solution: 50% ethanol solutions treated with pulsed electric fields for 1, 4, 7, 11 and 16 acceleration cycles. The pulsed electric field treatment of 50% ethanol solution was performed using a CorrTest electrochemical workstation with a two-electrode system. Glassy carbon electrodes were used as both the working electrode and counter electrode, and no reference electrode was adopted. The potential was scanned from 0.8 V down to -0.8 V gradually, and then returned to 0.8 V. A single scanning cycle lasted for 2 seconds, and the whole procedure was repeated 50 times. To monitor the temporal stability of the samples, 30 mL of sample was taken every 7 days for subsequent detection and analysis.

Instruments: Wuhan CorrTest Electrochemical Workstation (Wuhan, China), Hitachi F-7000 Fluorescence Spectrophotometer, Platinum Mesh Electrode (2 cm*2 cm), Calomel Electrode, Constant Temperature Water Bath Pot (Shanghai, China), Anton Paar DMA 4500M Precision Alcohol Density Meter, Sanliang Magnetic Stirrer (Guangdong, China).

### 2.2 Experimental Methods

#### 2.2.1 Electrochemical experiment

**2.2.1.1** Open circuit potential test. The lipid-polymer membrane is prepared by mixing dotriacontyl ammonium bromide, di-n-octylphenyl phosphonate, polyvinyl chloride and trimellitic acid in a certain proportion. PVC is blended with a certain amount of lipid and plasticizer in 10 mL of tetrahydrofuran under stirring for 1 hour, and the mixed solution is dried at room temperature for 72 hours to form a transparent sensitive membrane. The change in membrane potential is expressed as the potential difference between the sample solution and the reference solution (30 mM KCl, 0.3 mM tartaric acid). The preparation of lipid-polymer membranes, sensor tests and data processing methods are all carried out in accordance with standard procedures reported in literatures(Ye, Ai, Wu, Onodera, Ikezaki & Toko, 2022).

**2.2.1.2** Impedance test. Take 20 mL of the test solution and place it in an electrolytic cell with a diameter of 3.5 cm. Use platinum wire electrodes as the counter electrode and reference electrode, and a self-prepared lipid membrane electrode as the working electrode to conduct potentiostatic EIS measurements. Apply sinusoidal perturbation within the frequency range from 0.1 MHz to 0.01 Hz (including 0.001 Hz), with a logarithmic resolution of 10 data points per decade. To ensure thermodynamic equilibrium and data repeatability, the entire measuring device is placed in a water bath maintained at a constant temperature of 25.0 ± 0.1°C. Obtain the full-frequency scanning graph, determine the optimal frequency, and perform fixed-frequency time scanning. All samples are tested in triplicate.

**2.2.2 Steady-state fluorescence spectroscopy.**

The measurements were performed using an F-7000 steady-state fluorescence spectrophotometer (Hitachi High-Technologies Corporation, Japan) with an excitation wavelength of 240 nm. All operations were carried out at a constant temperature of 25°C and away from light to minimize photobleaching and thermal fluctuations. Transfer 2 mL of the sample into a quartz cuvette for scanning. After full-spectrum scanning, the fluorescence intensity values at the characteristic emission wavelengths (308, 330 and 370 nm) were extracted. All data are presented as the average values of three replicate measurements. Three replicate experiments were conducted for each sample to ensure statistical reliability.

**2.2.3 Determination of alcohol dehydrogenase activity.**

To accurately measure the proportion of free ethanol in ethanol-water binary systems, an enzymatic oxidation assay method based on alcohol dehydrogenase (ADH) was adopted**(Shang, Jiang, Zuo & Xie, 2026).** The total volume of the reaction system was 950 µL, consisting of 300 µL Tris-HCl buffer solution (3 mol/L, pH 8.8), 300 µL β-nicotinamide adenine dinucleotide ($NAD^+$) solution (2.5 mg/mL), 50 µL ethanol-water binary samples (volume fraction: 10%–50%), and 300 µL 100-fold diluted ADH stock solution (specific activity > 300 U/mg). The mixed solution was incubated in a constant temperature water bath at 37 °C for 96 hours to ensure the completion of the enzymatic reaction. Based on the characteristic fluorescence properties of the reaction product reduced nicotinamide adenine dinucleotide (NADH), steady-state fluorescence spectroscopy was used to quantify the reaction extent. A fluorescence spectrophotometer was applied for detection with the excitation wavelength set at 340 nm. According to the calibration curve established with standard ethanol solutions, the concentrations of free ethanol in each sample were calculated via the measured fluorescence intensity values.

**2.2.4** Sensory evaluation.

Six trained panelists were selected to conduct the sensory evaluation. Each panelist held the sample in the mouth, rinsed for 5 seconds and then swallowed it. The initial sweetness intensity (instant sweetness perceived within 0–3 seconds after sample intake, scored on a 1–10 scale) and lingering sweetness intensity (the sweetness perceived 10 seconds after swallowing, scored on a 1-10 scale, with the duration of the aftertaste recorded). After tasting each sample, panelists rinsed their mouths with ultrapure water and rested for 10 minutes before evaluating the next sample. Each panelist performed three repeated tests on each sample, resulting in a total of 18 tests (6 panelists × 3 repetitions). Origin 2024 software was used for statistical analysis. The results were expressed as mean ± standard deviation, and the valid sample size after eliminating outliers was n=12.

# 3. Result

## 3.1. Characterization of the lingering sweetness characteristics of ethanol clusters

To establish the correlation between ethanol cluster content and sensory characteristics, a series of samples with varying degrees of clustering were prepared via electric field accelerated aging(Jiang et al., 2026; Shang et al., 2026), and their structures were systematically characterized. Figure 2a presented the three-dimensional fluorescence spectra of 50% ethanol samples subjected to electric field accelerated treatment for different durations (1, 4, 7, 11, 16 weeks) at an excitation wavelength $\lambda_{ex}$=240 nm. Within the emission wavelength range of 300-400 nm, the samples showed three characteristic emission peaks(Jia, Li, Zhang, Gao & Wu, 2020) located at 308 nm, 338 nm and 370 nm respectively. Based on the empirical assignment of fluorescence spectra of ethanol-water systems in previous studies, these three peaks could be attributed to ethanol hydrated clusters with different degrees of hydrogen bond association. With the extension of electric field treatment time, the fluorescence intensities of the three characteristic peaks increased significantly in a gradient manner (Figure 2a), which indicated that electric field treatment effectively promoted the directional transformation of ethanol molecules from the free state to the clustered associated state, and the cluster content accumulated continuously with the increase of treatment time.To further quantify the ratio of free ethanol to associated ethanol in the system, ADH was adopted for differential detection. The active sites of ADH could only specifically recognize and catalyze the oxidation reaction of free ethanol, reducing $NAD^+$ to NADH with characteristic fluorescence(Shang et al., 2026). Figure 2b presented the enzymatic hydrolysis results of samples treated by electric field acceleration. The fluorescence intensity during the enzymatic reaction showed a monotonically decreasing trend.Figure S1 displayed the regression curve of fluorescence response during enzymatic hydrolysis for ethanol-water standard systems with ethanol concentrations from 10% to 50% (v/v). A significant positive linear correlation existed between fluorescence intensity and ethanol concentration ($y = 1.263x + 317.75$, $R^2 = 0.988$).After calculation, the relative contents of clusters in samples subjected to electric field acceleration for 1, 4, 7, 11 and 16 weeks were 4.83%, 9.82%, 15.99%, 23.67% and 29.53%, respectively. These quantitative data provided reliable physicochemical parameters for establishing the correlation between structure and sensory properties in subsequent sensory evaluation.

To explore the effect of ethanol cluster content on sweetness perception, standardized sensory evaluations were conducted on ethanol-water model system samples with varying cluster contents (4.83%-29.53%, without any added sweet substances) and xylitol solutions (50-250 mM) serving as controls. Figure 2c established a three-dimensional statistical distribution model of sensory sweetness evaluation. The perceived sweetness intensity of both xylitol and ethanol cluster samples was scored on a 1–10 scale, calibrated to 1–10% (w/v) sucrose solutions as reference standards, and was expressed as sucrose equivalent (%).The sweetness intensity of xylitol samples showed a significant positive linear correlation with increasing concentration (±0.8 points at 50 mM and ±3.2 points at 250 mM), which conformed to the classic concentration-response rule of sweet substances. Nevertheless, the duration of sweetness perception remained within a short time window of less than 10 seconds at all concentrations (Figure 2c), indicating that although xylitol could deliver intense instantaneous sweetness, it lacked flavor persistence and failed to produce the sensory experience of lingering sweetness. Samples containing ethanol clusters exhibited completely different sensory characteristics. Their overall sweetness intensity stayed at a medium to low level (1.0-1.7 points), while the duration of sweetness perception was markedly prolonged in a gradient manner with the rise of ethanol cluster content, increasing gradually from ±10 seconds at a cluster content of 4.83% to ±25 seconds at 29.53%, with a growth rate of over 150% (Figure 2c). The differences in perception duration among all groups were statistically significant (n=12, one-way ANOVA, $P<0.001$).

The above results indicated that there was a significant positive correlation between the content of ethanol clusters and the duration of sweet taste perception; the higher the cluster content, the longer the lasting time of lingering sweetness. This finding strongly supported the hypothesis that ethanol clusters played a crucial structural role in the sensory experience of lingering sweetness. Unlike traditional sweeteners such as xylitol, which enhanced sweet taste intensity by increasing concentration, ethanol clusters seemed to mainly act on the temporal dimension of sweet taste perception, achieving unique regulation of flavor persistence. This regulatory mechanism in the temporal dimension was to be further verified and analyzed through subsequent electronic tongue kinetic experiments.

In conclusion, sensory evaluation established a positive correlation between cluster content and lingering sweetness duration from the perspective of human perception, providing direct sensory evidence for the core finding of this study that ethanol clusters served as the material basis for the lingering sweetness effect.

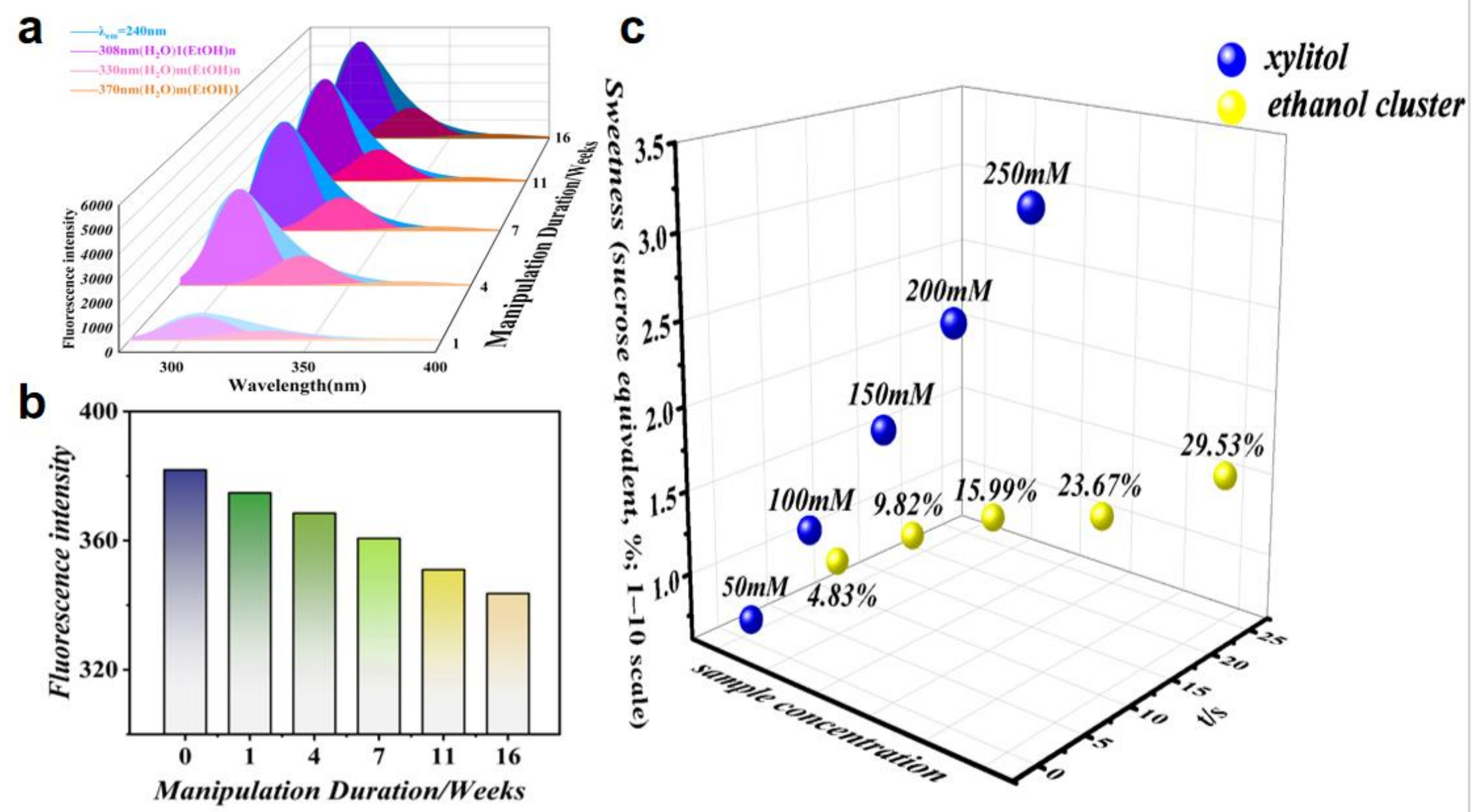

Figure 2 Changes of ethanol clusters, reduction of free ethanol and sensory sweetness evaluation during electric field accelerated aging. (a) Fluorescence emission spectra of samples at different accelerated aging times ($\lambda_{ex}$ = 240 nm); (b) Content variation of free ethanol during accelerated aging; (c) Three-dimensional scatter plot of sensory sweetness of xylitol (50–250 mM) and ethanol clusters (4.83%–29.53%) (expressed as sucrose equivalent, scored on a 1–10 scale),the y-axis represented normalized sample concentration (0–1 range) for both xylitol and ethanol clusters.

### 3.2. Performance Advantages of Impedance-Type Electronic Tongue

Traditional electronic tongue technologies (such as open-circuit potential method) mainly relied on the drift of membrane potential(Ye et al., 2022) at the lipid film interface to respond to changes in ionic strength in solutions (Wu et al., 2025)(Figure 3a1). However, key sweet-tasting substances including sucrose and ethanol clusters were uncharged molecules, which could hardly generate obvious potential differences on the surface of lipid films. This led to inherent drawbacks of traditional potentiometric methods such as low sensitivity and narrow linear range when detecting such substances. We made key improvements to the electronic tongue sensing principle proposed by Professor Toko Kiyoshi and introduced electrochemical EIS technology (Figure 3b1). This mechanism utilized the specific binding between sweet molecules and lipid film receptors to alter the dielectric constant and ion transport channels at the film interface, thereby causing a significant change in charge transfer resistance. The detection strategy based on interfacial impedance changes could theoretically break through the bottleneck in the detection of uncharged molecules and achieve higher detection sensitivity.

To verify the analytical performance of the impedance method, comparative experiments were first conducted using sucrose solutions of different concentrations (0–1000 mM) as model systems. When measured via the open-circuit potential method (Figure 3a2, c1), the response potential showed a slight upward trend with the increase of sucrose concentration, yet it exhibited poor linear fitting goodness. Linear regression analysis (Figure 3c1) revealed that the coefficient of determination ($R^2$) between potential response values and sucrose concentration was merely 0.424 with an extremely low slope (y=−0.0292x−12.179), which confirmed that traditional methods had obvious signal lag and nonlinear errors in the detection of uncharged sweet-tasting molecules. In contrast, the impedance method (Figure 3b2, c1) delivered outstanding analytical performance. Nyquist plots (Figure 3b2) indicated that the radius of the capacitive arc increased remarkably as the sucrose concentration rose, demonstrating that the accumulation of high-concentration sweet-tasting molecules at the membrane interface effectively hindered electron transfer. Figure 3c1 further verified that there was an extremely high linear correlation between impedance response values and sucrose concentration, with the $R^2$ reaching up to 0.998 and a greatly improved slope.

After confirming the excellent response of the impedance method to sucrose, it was further applied to detect ethanol-water system samples with different cluster contents ranging from 4.83% to 29.53%. As shown in Figure 3c2, the potentiometric method showed a certain linear trend with an $R^2$ value of 0.774 when detecting ethanol-water systems, yet it suffered from substantial data dispersion. In contrast, the impedance method achieved a closer linear relationship between the response signal and cluster content, with the $R^2$ value rising to 0.934. Consequently, comparative tests on sucrose model systems and ethanol-water system samples with varying cluster contents distinctly verified the improved linearity of the modified impedance-type electronic tongue over the traditional potentiometric method.

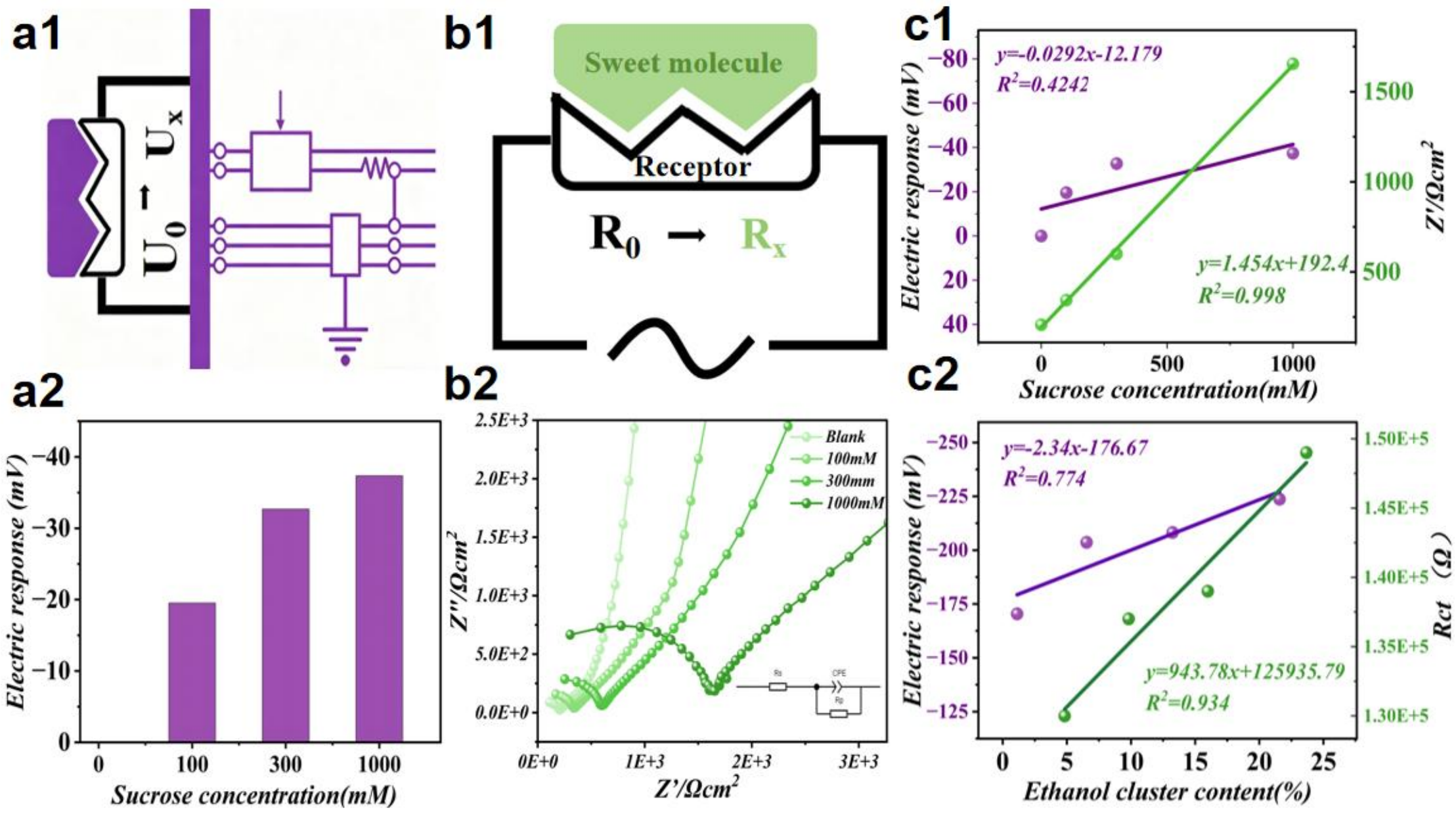

Figure 3 Comparison of responses to sweet substances and schematic diagram of sensing principles by open-circuit potential method and impedance method. (a1) Schematic diagram of the sensing principle of the open-circuit potential method; (a2) Potential responses of the open-circuit potential method to sucrose at different

concentrations (0–1000 mM); (b1) Schematic diagram of the sensing principle of the impedance method; (b2) Nyquist plots of the impedance method for sucrose at different concentrations (0–1000 mM); (c1) Linear response comparison between the open-circuit potential method (purple, $R^2$=0.424) and the impedance method (green, $R^2$=0.998) in the sucrose system; (c2) Linear response comparison between the open-circuit potential method (purple, $R^2$=0.774) and the impedance method (green, $R^2$=0.934) in the ethanol cluster system.

### 3.3. It is confirmed via electronic tongue analysis that ethanol molecular clusters possess sweet taste response activity.

As established in Section 3.2, sweet molecules binding to the lipid membrane receptor altered the interfacial dielectric constant and charge transfer resistance, which in turn modulated the impedance phase angle at different frequencies. Thus EIS combined with convolution analysis was employed. Figure 4a presented the phase angle-frequency relationship curves of xylitol samples processed by convolution after full-band scanning. The results indicated that the peak phase angle rose gradually as the xylitol concentration increased from 50 mM to 250 mM, demonstrating that higher concentrations could enhance the sweet taste signal response. By comparing the variations of phase angles at different frequencies, 0.933 Hz was identified as the optimal scanning frequency for xylitol samples.At this frequency, the phase angle showed the highest sensitivity to concentration changes, which could effectively distinguish the differences in sweet taste responses among various concentration gradients. Figure 4b showed the phase angle-frequency curves of ethanol cluster samples. The peak phase angle increased regularly with the growth of ethanol cluster content, which meant that higher content of ethanol clusters led to stronger sweet taste signals. Through full-band scanning analysis, 0.03 Hz was confirmed as the optimal scanning frequency for ethanol cluster samples. There was a remarkable difference in the optimal scanning frequencies between xylitol and ethanol clusters, which reflected the essential distinction in their molecular relaxation times(Bergmann & Schlüter, 2022). As a small-molecule sweetener, xylitol featured fast molecular movement and short relaxation time ($\tau \approx 0.171$ s), corresponding to the rapid response property of initial sweet taste. In contrast, ethanol clusters had relatively slow molecular movement and long relaxation time ($\tau \approx 5.305$ s) due to the stability of their supramolecular structures, which was consistent with the sustained release characteristic of lingering sweetness. The relaxation time extracted from EIS reflects the overall kinetics of molecular binding and dissociation at the lipid membrane interface, which is influenced by both molecular diffusion and conformational changes. According to the relationship between characteristic frequency (f) and relaxation time ($\tau$) in electrochemical EIS, $\tau=1/(2\pi f)$(Wang et al., 2026). For xylitol at 0.933 Hz, $\tau \approx 0.171$ s, and for ethanol clusters at 0.03 Hz, $\tau \approx 5.305$ s, which were exactly the values estimated from the phase-angle responses. This quantitative consistency confirmed that the observed frequency separation directly reflected the intrinsic kinetic difference between fast and slow molecular binding/dissociation processes.

Figure 4c further compared the variation rule of the phase angle of xylitol and ethanol clusters with their contents at the optimal scanning frequency. For xylitol (50-250 mM, 0.933 Hz), the phase angle showed a significant positive linear correlation with concentration ($y=0.0662x+52.918$, $R^2=0.941$), indicating that its sweetness signal intensity increased steadily with the rise of concentration, which was consistent with the dose-effect relationship of small-molecule sweeteners. Ethanol cluster samples (4.83%-29.53%, 0.03 Hz) also presented a linear increasing trend of phase angle with growing content ($y=1.034x+23.812$，R2=0.931), and the phase angle of ethanol clusters showed a steeper concentration dependence on a per-unit basis, reflecting different response characteristics of ethanol clusters. From the perspective of electrochemical impedance, this result verified that ethanol clusters could specifically bind to sweet taste receptors via their unique molecular structures to generate distinct sweetness signals, and the signal intensity was positively correlated with the cluster content.

Comprehensive analysis showed that the EIS convolution analysis method could effectively distinguish the response characteristics of different sweet substances. Xylitol generated an initial sweet taste signal through rapid molecular movement, while ethanol clusters achieved the sustained release of lingering sweet taste signals via stable molecular structures. The differences between the two in optimal scanning frequency, relaxation time and concentration response slope revealed the temporal mechanism of sweet taste perception from the perspective of molecular dynamics.

Thus, the impedance-based electronic tongue achieved, for the first time, an objective differentiation between upfront sweetness and Lingering Sweetness based on their distinct frequency responses and relaxation times. The high-frequency response of xylitol (0.933 Hz) and the low-frequency response of ethanol clusters (0.03 Hz) not only confirmed the existence of two sweetness modalities at the physical measurement level but also provided the first electronic-tongue-based objective criterion and kinetic interpretation framework for understanding the temporal dynamics of sweet taste perception through the quantitative difference in relaxation times.

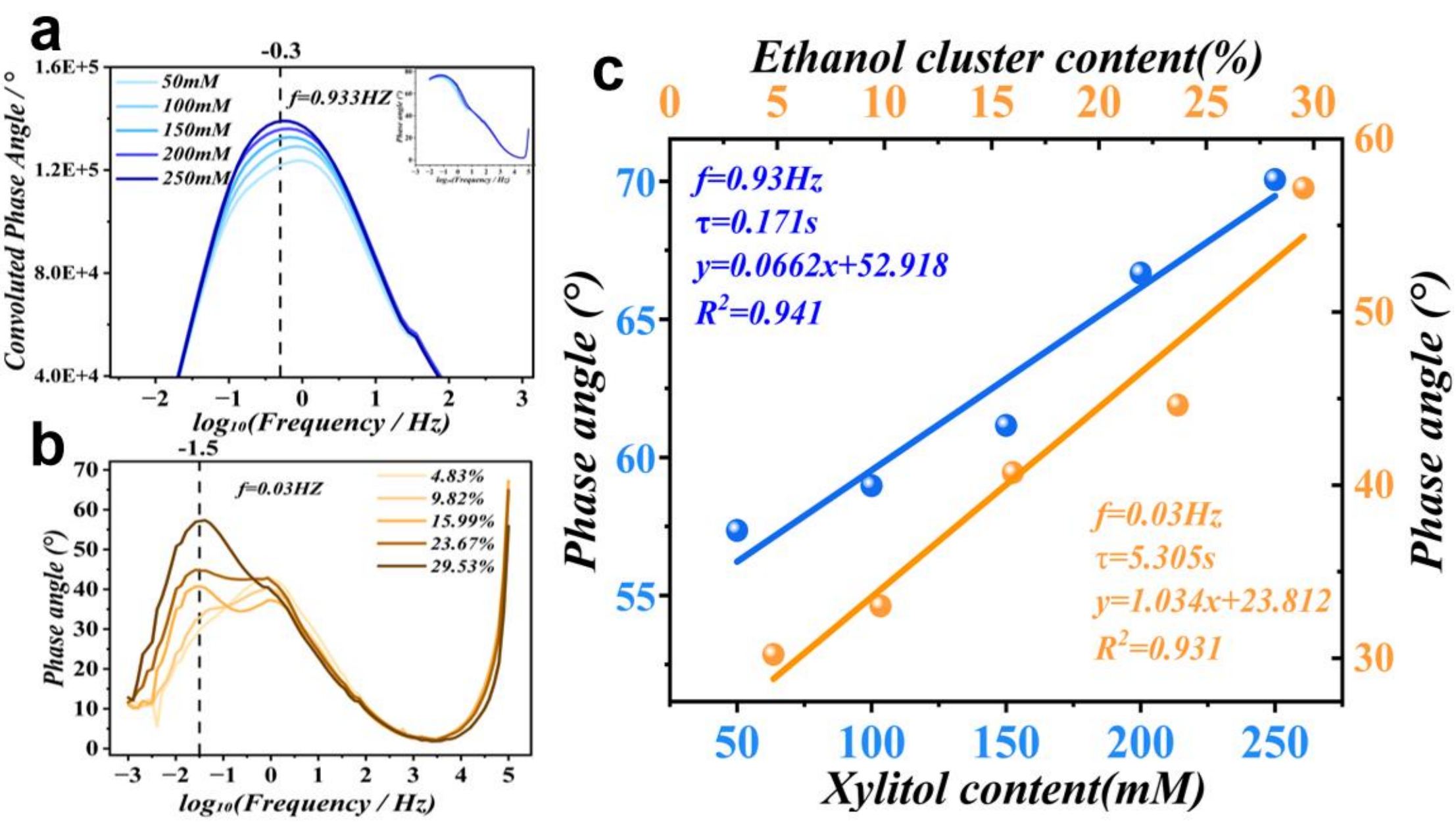

Figure 4 Impedance phase angle response and relaxation time characteristics of xylitol-ethanol clusters. (a) Full-frequency impedance phase angle convolution spectra of xylitol at different concentrations (50–250 mM) (the abscissa represents log10 frequency/Hz, and the ordinate represents the convolved phase angle intensity); (b) Full-frequency phase angle spectra of ethanol clusters with different contents (4.83%–29.53%); (c) Linear variations of phase angle responses of xylitol (blue, 0.933 Hz) and ethanol clusters (yellow, 0.03 Hz) with concentration and content.

### 3.4. Analysis of Sweetness Kinetic Response and Molecular Mechanism

This study explored the temporal dynamic characteristics of sweet substances and analyzed the phase angle responses measured by EIS and the sweetness duration obtained from sensory evaluation (Figure 5a, b). Figure 5a illustrated that with the increase of xylitol concentration (50–250 mM), the phase angle at the characteristic frequency (0.933 Hz) showed a significantly synchronous rising trend along with the sensory sweetness duration. This high consistency indicated that the increase in phase angle accurately reflected the rapid adsorption and dissociation kinetics of xylitol molecules at the receptor interface, which was macroscopically manifested as the initial sweet taste property. Figure 5b further revealed the unique dynamic behavior of ethanol cluster systems. As the content of ethanol clusters rose, the phase angle at the characteristic frequency was strongly positively correlated with the sensory sweetness duration. The sweetness duration of ethanol cluster systems was remarkably longer than that of xylitol systems, and it prolonged with the increase of cluster content. This result electrochemically confirmed that ethanol clusters acted not only as carriers of sweet taste signals but also as key regulators governing the temporal dimension of sweetness, and the rise in their content directly intensified the long aftertaste effect.

Based on the above experimental data, Figure 5c proposed a molecular mechanism model for the lingering sweetness effect induced by ethanol clusters. We hypothesize that xylitol had a small molecular volume, and its binding mode with sweet taste receptors was characterized by rapid binding and loose attachment. Such weak interactions led to rapid binding and dissociation of molecules within the receptor binding domain, thereby generating transient and short-lasting sweet taste perception at the macroscopic level. In contrast, ethanol clusters formed large supramolecular structures via hydrogen bond networks. When entering the receptor transmembrane domains or allosteric sites, these structures featured slow binding and tight anchoring. Larger cluster structures required more time for conformational adjustment to adapt to receptor pockets, and once bound, their multi-site hydrogen bond interactions markedly reduced the molecular dissociation rate.

This slow and tight binding kinetics manifested macroscopically as the persistent retention of sweet taste signals, namely a prominent lingering sweetness. In addition, with the increase in ethanol cluster content, their ordered arrangement at the receptor interface might further mitigate the irritating impact of free ethanol, making the overall taste become milder and mellower while the sweet taste lasted longer.

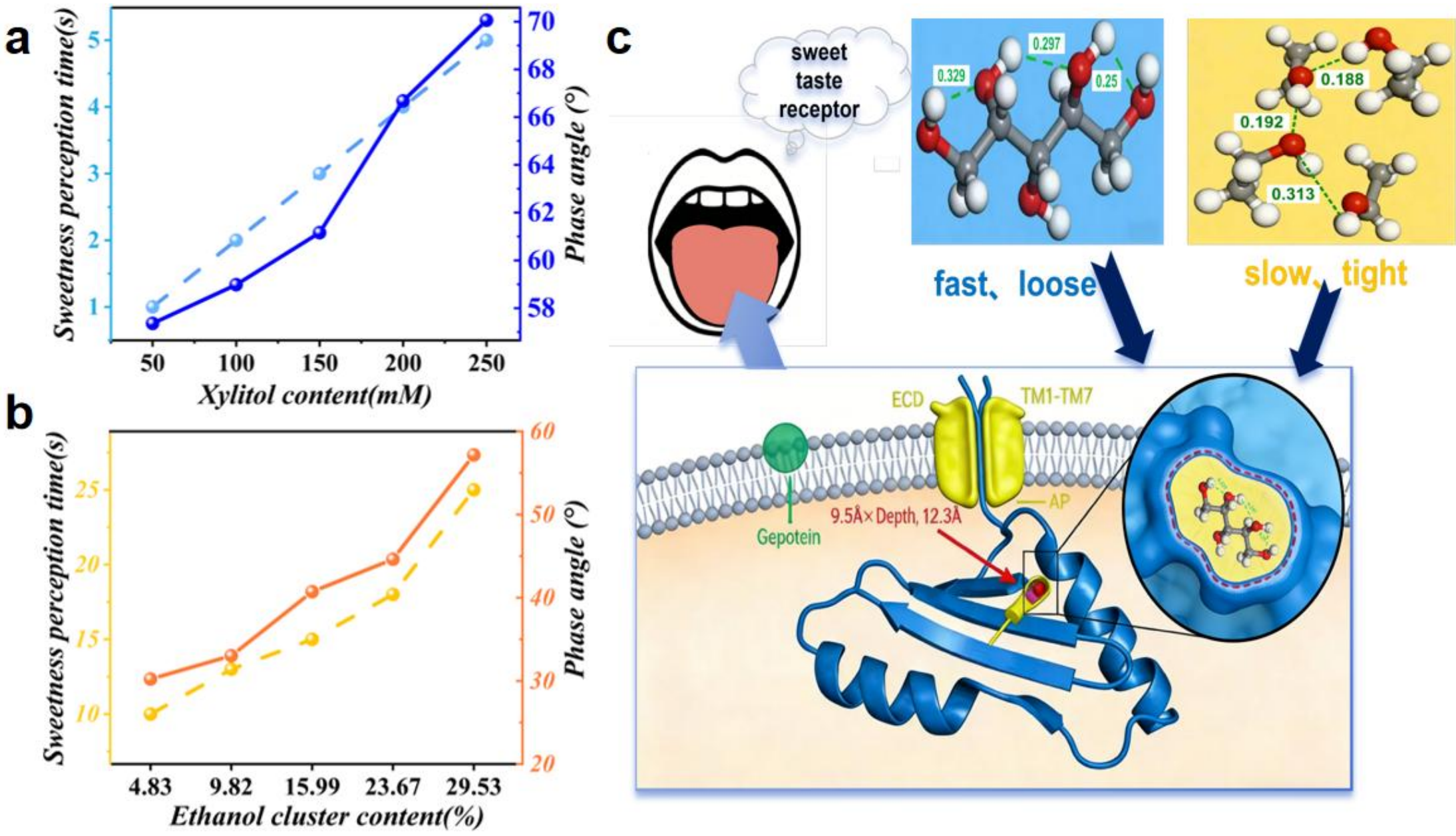

Figure 5 Schematic diagram of differences in sweet taste perception kinetics and receptor interaction mechanisms between xylitol and ethanol clusters. (a) Variation trends of sensory sweetness perception time (dashed line, light blue left axis) and impedance phase angle response (solid line, blue right axis) versus concentration (50–250 mM) in the xylitol system; (b) Variation trends of sensory sweetness perception time (dashed line, yellow left axis) and impedance phase angle response (solid line, orange right axis) versus content (4.83%–29.53%) in the ethanol cluster system; (c) Schematic diagram of the interaction mechanism of sweet taste receptors.

### 3.5 Correlation Analysis Between Sensory Lingering Sweetness Characteristics and EIS of Baijiu in Different Years

Full-band impedance scanning revealed that the optimal response frequency of natural aged Baijiu samples (0.01 Hz) was significantly lower than that of the model ethanol-water system (0.03 Hz). Figure S2 compared the phase angle-frequency profiles of a representative Baijiu sample (10-year aged) and the 29.53% ethanol cluster model system. The downward shift in optimal frequency indicated a longer relaxation time ($\tau$ = 15.92 s for Baijiu vs. 5.305 s for the model system). This difference was attributed to the higher micro-viscosity and more densely cross-linked hydrogen bond network(Qin et al., 2022) in real Baijiu, resulting from the presence of diverse flavor compounds (e.g., ethyl acetate, lactic acid, and higher alcohols)(Jiang et al., 2024). These matrix components not only stabilized ethanol clusters but also slowed their interfacial dynamics, thereby requiring a longer embedding time into the lipid membrane. This observation further supported the proposed mechanism that the lingering sweetness signal originated from the slow binding of ethanol clusters to sweet taste receptors.

Figure 6a presented the phase angle responses of liquor of different ages under low-frequency scanning at 0.01 Hz ($\tau$=15.92 s). The results showed that the phase angle exhibited a significant monotonically increasing trend as the aging time extended from 1 year to 20 years. According to literature reports, ethanol molecules gradually formed more stable and ordered cluster structures through hydrogen bonding during the aging process of Baijiu(Jiang et al., 2024). Based on the electrochemical impedance tests of ethanol clusters, the longer the aging years, the higher the content of ethanol clusters and the more prominent the sweet taste signal (aftertaste sweetness characteristics). Figure 6b was a sensory radar chart covering six dimensions: upfront sweetness intensity, lingering sweetness intensity, lingering sweetness duration, smoothness, irritation and refreshing degree. The sensory evaluation results were highly consistent with electronic tongue data. With the increase of aging years, the duration and intensity of aftertaste sweetness were significantly enhanced, which was in line with the growth rule of phase angle in electrochemical impedance results, indicating that the accumulation of ethanol clusters remarkably prolonged the perception duration of lingering sweetness. Meanwhile, the overall sensory profile shifted toward higher aftertaste sweetness, lower irritation and better smoothness. In contrast, there was little difference in initial sweetness intensity among different aging years, which proved that the instantaneous sweetness contributed by small-molecule sugars was not the dominant factor for the evolution of sweet taste during aging.

In conclusion, the phase angle signals detected by the electronic tongue were highly positively correlated with the duration of lingering sweetness in manual sensory evaluation. Together, they revealed that the evolution of ethanol cluster structures during the aging process served as the core mechanism for the enhancement of the lingering sweetness characteristic of Baijiu. This finding not only provided an objective and quantitative technical method for Baijiu quality evaluation, but also opened up a new approach to understanding the sweetness contribution mechanism of non-sugar substances in complex liquor bodies.

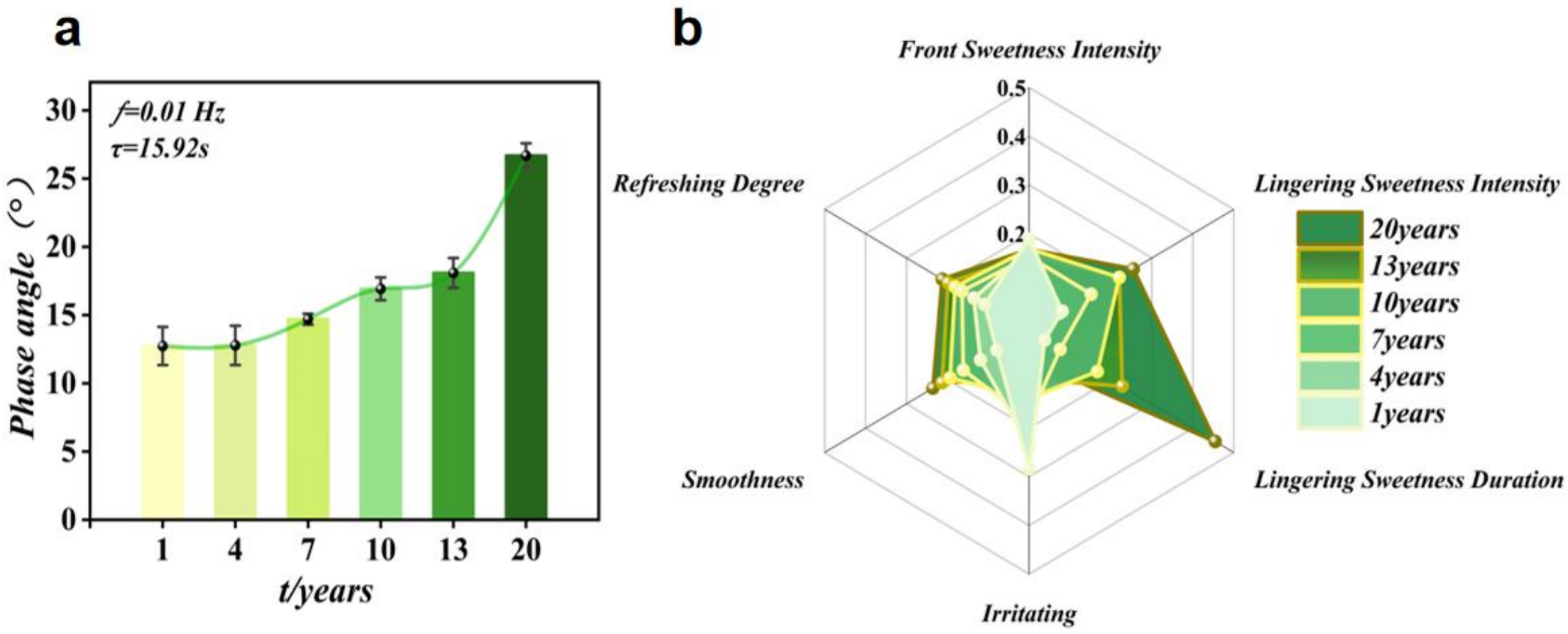

Figure 6 Phase angle response and sensory flavor characteristics of wine samples with different aging years. (a) Phase angle response of wine samples aged 1–20 years at the characteristic frequency (f=0.01 Hz, τ=15.92 s); (b) Sensory radar chart of wine samples aged 1–20 years.

## 4. Conclusion

This study firstly discovered the independent sweet taste function of ethanol clusters, and realized the objective distinction and kinetic characterization of upupfront sweetness and lingering sweetness via the innovative impedance-type electronic tongue technology.

(1)Ethanol clusters serve as the material basis for the lingering sweetness effect. A series of ethanol-water model systems with cluster contents ranging from 4.83% to 29.53% were prepared via electric field accelerated aging technology. Orthogonal verification through fluorescence spectroscopy and alcohol dehydrogenase enzymatic hydrolysis experiments confirmed the gradient variation of cluster contents. Sensory evaluation results indicate that the content of ethanol clusters has a significant positive correlation with the duration of sweet taste perception. The higher the cluster content, the longer the retention time of sweet taste (extended from 10 seconds to 25 seconds, $P < 0.001$), while the sweet taste intensity remains at a medium to low level consistently. This finding reveals that the contribution of ethanol clusters to sweet taste lies in the temporal dimension rather than the intensity dimension, which determines how long the sweetness lasts instead of how strong the sweetness is. It provides crucial experimental evidence for the expansion of the traditional AH-B sweet taste theory from single molecules to supramolecular clusters.

(2) For the first time, the impedance-type electronic tongue has realized the objective distinction between upupfront sweetness and lingering sweetness. After upgrading the traditional open-circuit potential method to the electrochemical EIS method, the coefficient of determination for sucrose detection increased from $R^2 = 0.424$ to $R^2 = 0.998$, the coefficient of determination for ethanol cluster detection increased from $R^2 = 0.774$ to $R^2 = 0.934$,which verifies the high-sensitivity response capability of this method to neutral sweet-tasting molecules. Full-band phase angle scanning and convolution analysis reveal that xylitol shows the strongest response in the high-frequency region of 0.933 Hz with a characteristic relaxation time of approximately 0.171 s, corresponding to the upupfront sweetness kinetics featuring rapid adsorption and rapid dissociation. Ethanol clusters achieve the optimal response in the low-frequency region of 0.03 Hz with a characteristic relaxation time of about 5.305 s, matching the lingering sweetness kinetics characterized by slow adsorption and slow dissociation. The strong correlation between phase angle response and sensory sweetness duration ($R^2 = 0.941$ for xylitol and $R^2 = 0.931$ for ethanol clusters) further confirms that high-frequency signals can characterize instant sweetness, while low-frequency signals specifically reflect lingering sweetness.

(3) Through the analysis of aged Baijiu samples with aging years ranging from 1 to 20 years, the findings of the experimental model were verified. As the aging period increases, the impedance phase angle signal rises monotonically. In sensory evaluation, the duration of lingering sweetness and smoothness improve simultaneously, while the pungency decreases. This confirms that in actual liquor bodies, over time, ethanol molecules form more stable and ordered cluster structures via hydrogen bond networks. Such structures not only neutralize the spicy and irritating sensation of free ethanol but also endow the liquor with a long-lasting lingering sweetness.

Based on the above findings, this study proposes a hypothesis on the molecular mechanism of ethanol cluster-induced lingering sweetness: small-molecule sweeteners such as xylitol bind to receptors rapidly and loosely, generating an instantaneous sweet taste. The supramolecular structure formed by ethanol clusters via hydrogen bond networks binds to receptors slowly and tightly, enabling the sustained retention of sweet taste signals. This mechanistic model connects macroscopic sensory experiences with microscopic molecular dynamics, offering a new theoretical framework for analyzing the temporal and spatial distribution of sweetness in complex systems. From a methodological perspective, the impedance-type electronic tongue technology adopted in this study establishes the first analytical platform for objective temporal evaluation in flavor science, and its analytical strategies based on characteristic frequency and relaxation time are expected to be extended to research on the temporal dynamics of other complex tastes including bitterness and astringency.